\documentclass[11pt,a4paper]{article}

\usepackage{amsmath,amsthm,latexsym,amssymb}
\usepackage{graphicx,bm,fancybox,multirow,hhline,indentfirst}
\usepackage{ascmac,atbegshi}
\usepackage{natbib,verbatim,color}
\usepackage[colorlinks,citecolor=blue,urlcolor=blue,dvipdfmx]{hyperref}

\makeatletter
\def\section{\@startsection {section}{1}{\z@}{-3.5ex plus -1ex minus -.2ex}{2.3 ex plus .2ex}{\large\sc\centering}}
\def\subsection{\@startsection {subsection}{1}{\z@}{-3.5ex plus -1ex minus -.2ex}{2.3 ex plus .2ex}{\large}}
\makeatother

\newtheorem{theo}{Theorem}
\newtheorem{coro}{Corollary}
\newtheorem{algo}{Algorithm}

\title{\Large\bf Use of spurious correlation for multiplicity adjustment \medskip}

\author{Yoshiyuki Ninomiya\thanks{744 Motooka, Fukuoka 819-0395, Japan. Email: nino@imi.kyushu-u.ac.jp}\\{\it Institute of Mathematics for Industry, Kyushu University} \medskip \and Satoshi Kuriki\thanks{10-3 Midori-cho, Tachikawa, Tokyo 190-8562, Japan. Email: kuriki@ism.ac.jp}\\{\it Department of Mathematical Analysis and Statistical Inference}\\{\it The Institute of Statistical Mathematics} \medskip \and Toshihiko Shiroishi\thanks{1111 Yata, Mishima, Shizuoka 411-8540, Japan. Email: tshirois@nig.ac.jp}\\{\it Mammalian Genetics Laboratory, National Institute of Genetics} \medskip \and Toyoyuki Takada\thanks{1111 Yata, Mishima, Shizuoka 411-8540, Japan. Email: ttakada@nig.ac.jp}\\{\it Mammalian Genetics Laboratory, National Institute of Genetics} \medskip}

\date{\normalsize Version: December 14, 2016}

\selectfont

\allowdisplaybreaks

\def\maxt{\textsc{maxt}}

\def\T{{\mathrm{\scriptscriptstyle T}}}
\def\E{E}
\def\V{\rm var}
\def\P{{\rm pr}}
\def\cor{{\rm cor}}

\begin{document}

\maketitle

\noindent Abstract:
We consider one of the most basic multiple testing problems that compares expectations of multivariate data among several groups. 
 As a test statistic, a conventional (approximate) $t$-statistic is considered, and we determine its rejection region using a common rejection limit. 
 When there are unknown correlations among test statistics, the multiplicity adjusted $p$-values are dependent on the unknown correlations. They are usually replaced with their estimates that are always consistent under any hypothesis. 
 In this paper, we propose the use of estimates, which are not necessarily consistent and are referred to as spurious correlations, in order to improve statistical power. 
 Through simulation studies, we verify that the proposed method asymptotically controls the family-wise error rate and clearly provides higher statistical power than existing methods.
 In addition, the proposed and existing methods are applied to a real multiple testing problem that compares quantitative traits among groups of mice and the results are compared.

\



\section{Introduction}\label{sec1}

We consider a simple multiple testing problem that compares the expectations of two-dimensional independent data from control and case groups.
 Setting the sample size as $10$ in each group, we denote the data in the control group by $\{(y_{i1}^{(0)},y_{i2}^{(0)})^{\T}\ |\ 1\le i\le 10\}$ and the data in the case group by $\{(y_{i1}^{(1)},y_{i2}^{(1)})^{\T}\ |\ 1\le i\le 10\}$.
 In addition, their expectations and variances are denoted by
\begin{align}
\E\left(\begin{array}{c}y_{i1}^{(u)}\\y_{i2}^{(u)}\end{array}\right)
=\left(\begin{array}{c}\mu_{1}^{(u)}\\\mu_{2}^{(u)}\end{array}\right),\qquad
\V\left(\begin{array}{c}y_{i1}^{(u)}\\y_{i2}^{(u)}\end{array}\right)
=\left( \begin{array}{cc} \sigma_{11}&\sigma_{12}\\\sigma_{12}&\sigma_{22}\end{array} \right)
\label{simplemodel}
\end{align}
($u\in\{0,1\}$), and we assume that they are unknown.
 For this model, we consider testing $H_1^{(1)}: \mu_1^{(1)}=\mu_1^{(0)}$ against $K_1^{(1)}: \mu_1^{(1)}>\mu_1^{(0)}$ and $H_2^{(1)}: \mu_2^{(1)}=\mu_2^{(0)}$ against $K_2^{(1)}: \mu_2^{(1)}>\mu_2^{(0)}$, simultaneously.
 As test statistics, we use conventional $t$-statistics, $T_1^{(1)}$ and $T_2^{(1)}$, and a rejection region is determined using a common rejection limit $c$.
 If the values of $(\sigma_{11},\sigma_{12},\sigma_{22})$ are known, then as a rejection limit, we have to only obtain the value of $c$ such that the family-wise error rate
\begin{align*}
\P_{H_1^{(1)}\cap H_2^{(1)}}\{\max (T_1^{(1)},T_2^{(1)})>c\}
\end{align*}
is controlled; that is, the family-wise error rate is equal to $\alpha$, where $\P_H$ refers to a probability under a hypothesis $H$ and $\alpha$ is the significance level for this multiple testing.
 It is to be noted that the higher the value of the correlation $\sigma_{12}/\sqrt{\sigma_{11}\sigma_{22}}$ is, the higher the correlation between $T_1^{(1)}$ and $T_2^{(1)}$ is. This would result in more number of tests being rejected.
 In this problem, the value of $\sigma_{jk}$ is unknown, and we intend to asymptotically control the family-wise error rate.

A natural choice would be to replace $\sigma_{jk}$ with its reasonable estimator such as an unbiased estimator 
\begin{align*}
\tilde{\sigma}_{jk}=\frac{1}{18}\sum_{u=0}^1\sum_{i=1}^{10}(y_{ij}^{(u)}-\bar{y}_j^{(u)})(y_{ik}^{(u)}-\bar{y}_k^{(u)})
\end{align*}
($j,k\in\{1,2\}$) in the asymptotic null distribution of $(T_1^{(1)},T_2^{(1)})^{\T}$, that is a two-dimensional Gaussian distribution with a mean of $0$, a variance of $1$, and a correlation of $\sigma_{12}/\sqrt{\sigma_{11}\sigma_{22}}$.
 We call this the \maxt\ method (see Section 2.6 of \citealt{DudLaa07}). 
 Because $\tilde{\sigma}_{jk}$ is consistent without relation to what hypothesis is true, this method asymptotically controls the family-wise error rate.

On the other hand, to improve statistical power, we evaluate the correlation by assuming that the expectations are the same in both groups.
 This means that we use
\begin{align*}
\hat{\sigma}_{jk} = \frac{1}{19} \sum_{u=0}^1 \sum_{i=1}^{10} \bigg( y_{ij}^{(u)} - \frac{\bar{y}_j^{(0)}+\bar{y}_j^{(1)}}{2} \bigg) \bigg( y_{ik}^{(u)} - \frac{\bar{y}_k^{(0)}+\bar{y}_k^{(1)}}{2} \bigg),
\end{align*}
an unbiased estimator of $\sigma_{jk}$ when $H_1^{(1)}\cap H_2^{(1)}$ is true, in place of $\tilde{\sigma}_{jk}$.
 The reason for this is the fact that $\hat{\sigma}_{12}/\sqrt{\hat{\sigma}_{11}\hat{\sigma}_{22}}$, the so-called spurious correlation, tends to be larger than $\tilde{\sigma}_{12} / \sqrt{\tilde{\sigma}_{11}\tilde{\sigma}_{22}}$ when an alternative hypothesis is true.
 In general, using such a spurious correlation does not assure any control of the family-wise error rate because it becomes a meaningless value under certain hypotheses; however, this spurious correlation does assure that the family-wise error rate is asymptotically controlled in this problem and can be seen in the next paragraph.

We will verify the asymptotic control in the following four cases: (a) when $H_1^{(1)}\cap H_2^{(1)}$ is true, (b) when $H_1^{(1)}\cap K_2^{(1)}$ is true, (c) when $K_1^{(1)}\cap H_2^{(1)}$ is true, and (d) when $K_1^{(1)}\cap K_2^{(1)}$ is true.
 In this method, we use $c$, such that $\P\{\max (\hat{X}_1^{(1)},\hat{X}_2^{(1)})>c\}=\alpha$, as a rejection limit for each test, where $(\hat{X}_1^{(1)},\hat{X}_2^{(1)})^{\T}$ is a two-dimensional Gaussian random vector with the mean being $0$, the variance being $1$ and the correlation being $\hat{\sigma}_{12}/\sqrt{\hat{\sigma}_{11}\hat{\sigma}_{22}}$.
 For case (a), since the spurious correlation is a consistent estimator, the family-wise error rate is evaluated as
\begin{align*}
\P_{H_1^{(1)}\cap H_2^{(1)}}\{\max (T_1^{(1)},T_2^{(1)})>c\}
\approx\P\{\max (\hat{X}_1^{(1)},\hat{X}_2^{(1)})>c\}=\alpha.
\end{align*}
 For case (b), although the spurious correlation is not consistent, it does not appear in the expression of the family-wise error rate. The family-wise error rate is evaluated as
\begin{align*}
\P_{H_1^{(1)}}(T_1^{(1)}>c)\approx\P(\hat{X}_1^{(1)}>c)\le\alpha.
\end{align*}
 For case (c), the spurious correlation does not appear in the expression of the family-wise error rate, which is similar to case (b). For case (d), the family-wise error rate is always zero.
 Therefore, we have verified the control.

\begin{figure}
\begin{center}
\includegraphics[scale=0.5]{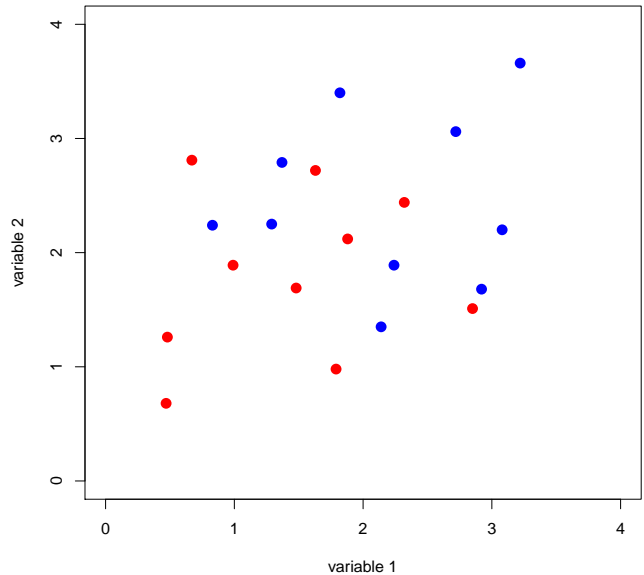}
\end{center}
\caption{Two-dimensional synthetic data from two groups. Correlations for control and case groups (red and blue) are $0.18$ and $0.12$, respectively, and the spurious correlation is $0.30$.}
\label{SimpleEx}
\end{figure}

Figure \ref{SimpleEx} plots synthetic data under $K_1^{(1)} \cap K_2^{(1)}$ in the setting above.
 Under $K_1^{(1)} \cap K_2^{(1)}$, the first and second variables in the case group are larger than those in the control group. Consequently, the spurious correlation becomes larger than the correlations in the two groups.
 For this data, we consider a multiple test consisting of the above-mentioned two tests, testing $H_1^{(1)}$ against $K_1^{(1)}$ and testing $H_2^{(1)}$ against $K_2^{(1)}$, with a significance level of $5\%$.
 While the \maxt\ method does not reject either of the tests, both tests are rejected when the spurious correlation is used.

It is not true that we can always use the spurious correlation.
 We assume one more case group which consists of independent data $\{(y_{i1}^{(2)},y_{i2}^{(2)})^{\T}\ |\ 1\le i\le 10\}$ satisfying \eqref{simplemodel} with $u=2$.
 We consider testing $H_1^{(2)}: \mu_1^{(2)}=\mu_1^{(0)}$ against $K_1^{(2)}: \mu_1^{(2)}>\mu_1^{(0)}$ and testing $H_2^{(2)}: \mu_2^{(2)}=\mu_2^{(0)}$ against $K_2^{(2)}: \mu_2^{(2)}>\mu_2^{(0)}$ in addition to the above-mentioned two tests, and we denote their test statistics by $T_1^{(2)}$ and $T_2^{(2)}$.
We then consider a four-dimensional Gaussian distribution which is the asymptotic null distribution of $(T_1^{(1)},T_2^{(1)},T_1^{(2)},T_2^{(2)})^{\T}$.
 Let $(\hat{\hat{X}}_1^{(1)},\hat{\hat{X}}_2^{(1)},\hat{\hat{X}}_1^{(2)},\hat{\hat{X}}_2^{(2)})^{\T}$ be a random vector distributed according to a distribution made by replacing $\sigma_{jk}$ with 
\begin{align}
\hat{\hat{\sigma}}_{jk} = \frac{1}{29} \sum_{u=0}^2 \sum_{i=1}^{10} \bigg( y_{ij}^{(u)} - \frac{\bar{y}_j^{(0)}+\bar{y}_j^{(1)}+\bar{y}_j^{(2)}}{3} \bigg)
\bigg( y_{ik}^{(u)} - \frac{\bar{y}_k^{(0)}+\bar{y}_k^{(1)}+\bar{y}_k^{(2)}}{3} \bigg),
\label{spur0}
\end{align}
which is an unbiased estimator under $H_1^{(1)}\cap H_1^{(2)}\cap H_2^{(1)}\cap H_2^{(2)}$ in the four-dimensional Gaussian distribution.
 We assume the value of $c$ such that $\P\{\max (\hat{\hat{X}}_1^{(1)},\hat{\hat{X}}_2^{(1)},\hat{\hat{X}}_1^{(2)},\hat{\hat{X}}_2^{(2)})>c\}=\alpha$ is the rejection limit of each test.
 Under this condition, we consider the family-wise error rate when $H_1^{(1)}\cap H_2^{(1)}\cap K_1^{(2)}\cap K_2^{(2)}$ is true.
If it holds for similar type of evaluations as in the case of the two groups, the family-wise error rate is expressed as
\begin{align}
\P_{H_1^{(1)}\cap H_2^{(1)}}\{\max (T_1^{(1)},T_2^{(1)})>c\}
\approx\P\{\max (\hat{\hat{X}}_1^{(1)},\hat{\hat{X}}_2^{(1)})>c\}\le\alpha;
\label{fwerex}
\end{align}
however, this approximation does not necessarily hold.
This is because of the fact that the approximation replaces $\sigma_{jk}$ with $\hat{\hat{\sigma}}_{jk}$ despite $\hat{\hat{\sigma}}_{jk}$ not being a consistent estimator of $\sigma_{jk}$ under this hypothesis.

In general, using a spurious correlation does not assure any asymptotic control of the family-wise error rate; however, for the correlation between two test statistics, using an estimator that is consistent under the null hypotheses in the two tests assures asymptotic control of the family-wise error rate.
 In \eqref{fwerex}, if the correlation between $T^{(1)}_1$ and $T^{(2)}_2$ is replaced with an estimator under $H^{(1)}_1\cap H^{(2)}_2$, the approximation holds.
 This theory is given in its general form in Section \ref{sec2}.

For correlated multiple tests without any pre-specified hypothesis ordering such as the above example, the \maxt\ method is conventional, and \cite{WesYou93} showed that it asymptotically controls the family-wise error rate under a subset pivotality condition.
 It was also shown by \cite{PolLaa04} and \cite{DudLP04} that the condition is relaxed by an easy algorithm.
 Moreover, the \maxt\ method can also be used in a step-down procedure (\citealt{LaaDP04}). 
 In this paper, under the same situation, we consider a different method that enhances statistical power.

In recent years, multiple testing procedures have been developed due to a rising demand from applications in the fields of medicine, bioinformatics, genomics, and brain imaging (see, e.g., \citealt{Far08}). 
 A well-developed approach that does not consider the use of correlations is known as the ``oracle approach''. This approach constructs an optimal test function by assuming that the true values of the parameters or their prior distributions are known.
 When the subject of what we want to control is the false discovery rate, the oracle approach works well if we substitute simple estimators for the true values of the parameters, or even if the prior distributions are slightly misspecified (\citealt{GenRW06}, \citealt{Sto07}, \citealt{SunCai07}, \citealt{GuiMZ09}). 
 On the other hand, it is difficult to control the family-wise error rate if we only use a simple estimator, and as shown in \cite{RoeWas09}, a natural choice would be to use a two-stage method with a sample-splitting procedure (\citealt{RubDL06}, \citealt{WasRoe06}, \citealt{HabPen14}). 
 In this method, the parameters are estimated from one split sample and testing is implemented by the other split samples.
 Although this assures the asymptotic control of the family-wise error rate, we have no appropriate theory on the splitting of the samples, and so an arbitrariness for the splitting exists.
 \cite{RoeWas09} avoided this two-stage method and roughly estimated the parameters on purpose in order to approximately control the family-wise error rate; however, it is still difficult to construct a theory on how to roughly estimate the parameters.
 The approach for improving statistical power using such methods is different from the one proposed in this paper, and it is an attractive future theme that combines the two approaches.


\section{General theory}\label{sec2}

Supposing that the covariance matrices for the groups are different from each other and the alternative hypotheses are two-sided, we generalize the method in the previous section.
 For simple notations, we study pairwise comparisons between one control group and multiple case groups but we can also study general pairwise comparisons (see Web Appendix).

Let us denote a $p$-dimensional random vector for the $i$-th sample in the $u$-th group by $(y_{i1}^{(u)},\ldots,y_{ip}^{(u)})^{\T}$ ($1\le i\le n^{(u)}$, $0\le u\le m$), and its mean vector and covariance matrix by $(\mu_1^{(u)},\ldots,\mu_p^{(u)})^{\T}$ and $\bm{\Sigma}^{(u)}=(\sigma_{jk}^{(u)})_{1\le j,k\le p}$, where $u=0$ and $u=1,\ldots,m$ indicate the control and case groups, respectively.
 We assume that the parameters of interest are in the mean vector and consider a multiple testing problem that compares $H_j^{(s)}: \mu_j^{(s)} = \mu_j^{(0)}$ and $K_j^{(s)}: \mu_j^{(s)}\neq\mu_j^{(0)}$ ($1\le j\le p,\ 1\le s\le m$). 
 By denoting the average of $y_{ij}^{(s)}$ by $\bar{y}_j^{(s)} = \sum^n_{i=1}y_{ij}^{(s)}/n^{(s)}$ and a conventional unbiased estimator of $\sigma_{jk}^{(s)}$ by $\tilde{\sigma}_{jk}^{(s)} = \sum_{i=1}^{n^{(s)}}(y_{ij}^{(s)}-\bar{y}_j^{(s)})(y_{ik}^{(s)}-\bar{y}_k^{(s)})/(n^{(s)}-1)$, an approximate $t$-statistic for each test is written by
\begin{align*}
T_j^{(s)} \equiv (\tilde{\sigma}_{jj}^{(s)} / n^{(s)} + \tilde{\sigma}_{jj}^{(0)} / n^{(0)})^{-1/2}(\bar{y}_j^{(s)}-\bar{y}_j^{(0)}).
\end{align*}
For this problem, when a common rejection limit $c$ is used in every test, the family-wise error rate is given by
\begin{align}
\P_{\bigcap_{1\le j\le p,1\le s\le m}H_j^{(s)}} \bigg( \max_{1\le j\le p,1\le s\le m}|T_j^{(s)}|>c \bigg)
\label{adjustedp}
\end{align}
under the complete null hypothesis.
 We can easily verify the method that uses the rejection limit $c$, such that the value in \eqref{adjustedp} is a significance level $\alpha$, keeps the family-wise error rate below $\alpha$ under any hypothesis; that is, it strongly controls the family-wise error rate.
 On the other hand, we cannot obtain an exact value of the tail probability because the distribution of $\max_{1\le j\le p,1\le s\le m}|T_j^{(s)}|$ depends on unknown parameters $\{\sigma_{jk}^{(s)}\ |\ 1\le j,k\le p,\ 1\le s\le m\}$ and we do not assume any parametric model for the distribution of data in the first place.

The Bonferroni method considers $\sum_{1\le j\le p,1\le s\le m} \P_{H_j^{(s)}} (|T_j^{(s)}|>c)$ as an upper bound of \eqref{adjustedp}, and as a rejection limit, it uses $c$ such that the value of the bound is $\alpha$; however, it is too conservative when the correlations among $\{T_j^{(s)}\ |\ 1\le j\le p,\ 1\le s\le m\}$ are large.
 In such a case, a natural choice would be to use an asymptotic evaluation of \eqref{adjustedp} (see Section 2.6 of \citealt{DudLaa07}). 
 Under the hypothesis $\bigcap_{1\le j\le p,1\le s\le m}H_j^{(s)}$, the asymptotic distributions of $T_j^{(s)}$'s are ${\rm N}(0,1)$ and their correlations $\cor (T_j^{(s)},T_k^{(s)})$, $\cor (T_j^{(s)},T_k^{(t)})$, $\cor (T_j^{(s)},T_j^{(t)})$ are asymptotically equivalent to
\begin{align}
& \rho_{jk}^{(ss)} \equiv (\tilde{\sigma}_{jj}^{(s)}/n^{(s)} + \tilde{\sigma}_{jj}^{(0)}/n^{(0)})^{-1/2}
(\tilde{\sigma}_{kk}^{(s)}/n^{(s)} + \tilde{\sigma}_{kk}^{(0)}/n^{(0)})^{-1/2}
(\tilde{\sigma}_{jk}^{(s)}/n^{(s)} + \tilde{\sigma}_{jk}^{(0)}/n^{(0)}),
\label{asympcor1}
\\
& \rho_{jk}^{(st)} \equiv (\tilde{\sigma}_{jj}^{(s)}/n^{(s)} + \tilde{\sigma}_{jj}^{(0)}/n^{(0)})^{-1/2}
(\tilde{\sigma}_{kk}^{(t)}/n^{(t)} + \tilde{\sigma}_{kk}^{(0)}/n^{(0)})^{-1/2}
\tilde{\sigma}_{jk}^{(0)}/n^{(0)},
\label{asympcor2}
\\
& \rho_{jj}^{(st)} \equiv (\tilde{\sigma}_{jj}^{(s)}/n^{(s)} + \tilde{\sigma}_{jj}^{(0)}/n^{(0)})^{-1/2}
(\tilde{\sigma}_{jj}^{(t)}/n^{(t)} + \tilde{\sigma}_{jj}^{(0)}/n^{(0)})^{-1/2}
\tilde{\sigma}_{jj}^{(0)}/n^{(0)},
\label{asympcor3}
\end{align}
respectively, where $\tilde{\sigma}_{jk}^{(u)}$ is a conventional unbiased estimator of $\sigma_{jk}^{(u)}$.
 Here, and hereafter, we assume $j\neq k$ and $s\neq t$.
 Let $\{\tilde{X}_j^{(s)}\ |\ 1\linebreak[0]\le j\le p,\ 1\le s\le m\}$ be a set of standard Gaussian variables whose correlations $\cor (\tilde{X}_j^{(s)},\tilde{X}_k^{(s)})$, $\cor (\tilde{X}_j^{(s)},\tilde{X}_k^{(t)})$, and $\cor (\tilde{X}_j^{(s)},\tilde{X}_j^{(t)})$ are given by $\rho_{jk}^{(ss)}$, $\rho_{jk}^{(st)}$, and $\rho_{jj}^{(st)}$, respectively.
 Then $\P(\max_{1\le j\le p,1\le s\le m}|\tilde{X}_j^{(s)}|>c)$ is asymptotically equivalent to \eqref{adjustedp}, and the method that uses $c$ such that its value is $\alpha$ asymptotically controls the family-wise error rate. We call this the \maxt\ method.

In this paper, we put `` $\hat{\ }$ (a hat) '' on estimators under a null hypothesis and on random variables based on the estimators while we put `` $\tilde{\ }$ (a tilde) '' on estimators that are always consistent without relation to what the true hypothesis is. 
 For an appropriate degree of freedom $d_{s}^{(u)}$, let us define
\begin{align}
\hat{\sigma}_{\{s\},jk}^{(u)} \equiv 
\frac{1}{n^{(u)}-d_{s}^{(u)}}\sum_{i=1}^{n^{(u)}} \bigg(y_{ij}^{(u)}-\frac{\sum_{v\in\{0,s\}} \sum_{i=1}^{n^{(v)}} y_{ij}^{(v)}}{\sum_{v\in\{0,s\}}n^{(v)}}\bigg) \bigg(y_{ik}^{(u)}-\frac{\sum_{v\in\{0,s\}} \sum_{i=1}^{n^{(v)}} y_{ik}^{(v)}}{\sum_{v\in\{0,s\}}n^{(v)}}\bigg)
\label{spur}
\end{align}
($u\in\{0,s\}$).
 Then, $(\hat{\sigma}_{\{s\},jk}^{(0)},\hat{\sigma}_{\{s\},jk}^{(s)})$ is a reasonable estimator of $(\sigma_{jk}^{(0)},\sigma_{jk}^{(s)})$ under a null hypothesis $H_j^{(s)}\cap H_k^{(s)}$.
 If these estimators are expected to be larger than $(\tilde{\sigma}_{jk}^{(0)},\tilde{\sigma}_{jk}^{(s)})$, we would want to use them in place of $(\tilde{\sigma}_{jk}^{(0)},\tilde{\sigma}_{jk}^{(s)})$ in order to obtain improved statistical power.
 Therefore, as a natural choice, we consider a method that uses $(\hat{\sigma}_{\{s\},jk}^{(0)},\hat{\sigma}_{\{s\},jk}^{(s)})$ in place of $(\tilde{\sigma}_{jk}^{(0)},\tilde{\sigma}_{jk}^{(s)})$ in the definition of $\rho_{jk}^{(ss)}$.
 That is, we define $\{\hat{X}_{j}^{(s)}\ |\ 1\le j\le p,\ 1\le s\le m\}$ as a set of standard Gaussian variables whose correlations $\cor (\hat{X}_{j}^{(s)},\hat{X}_{k}^{(s)})$ are given by using the replacement in the definition of $\rho_{jk}^{(ss)}$, and the correlations $\cor (\hat{X}_{j}^{(s)},\hat{X}_{k}^{(t)})$ and $\cor (\hat{X}_{j}^{(s)},\hat{X}_{j}^{(t)})$ are given by $\rho_{jk}^{(st)}$ and $\rho_{jj}^{(st)}$, respectively, and then we consider a method that uses $c$, such that $\P (\max_{1\le j\le p,1\le s\le m}|\hat{X}_{j}^{(s)}|>c)=\alpha$, as a common rejection limit.
 If the correlation matrix obtained in this manner does not satisfy the positive-definiteness, we will make it approach the conventional correlation matrix so that it satisfies the positive-definiteness, and we will then use the approached matrix.
 We call this the \maxt\ method using $(\hat{\sigma}_{\{s\},jk}^{(0)},\hat{\sigma}_{\{s\},jk}^{(s)})$ for $\rho_{jk}^{(ss)}$.
 Although using estimators under a null hypothesis does not assure any control of the family-wise error rate in general, the \maxt\ method using $(\hat{\sigma}_{\{s\},jk}^{(0)},\hat{\sigma}_{\{s\},jk}^{(s)})$ for $\rho_{jk}^{(ss)}$ assures the control of the family-wise error rate.
 This assurance is written in a general form as follows
 (See Appendix for its proof):

\begin{theo}
The \maxt\ method using consistent estimators of $(\sigma_{jk}^{(0)},\sigma_{jk}^{(s)})$ under $H_j^{(s)}\cap H_k^{(s)}$ in place of $(\tilde{\sigma}_{jk}^{(0)},\tilde{\sigma}_{jk}^{(s)})$ for $\rho_{jk}^{(ss)}$ in \eqref{asympcor1} asymptotically controls the family-wise error rate.
\label{th1}
\end{theo}

Conversely, the \maxt\ method using inconsistent estimators of $(\sigma_{jk}^{(0)},\sigma_{jk}^{(s)})$ under $H_j^{(s)}\cap H_k^{(s)}$ in place of $(\tilde{\sigma}_{jk}^{(0)},\tilde{\sigma}_{jk}^{(s)})$ for $\rho_{jk}^{(ss)}$ does not necessarily control the family-wise error rate.
 Let us again consider the example of the three groups in Section \ref{sec1}. 
 The fact that $\hat{\hat{\sigma}}_{jk}$ in \eqref{spur0} does not necessarily have a consistency under $H_j^{(1)} \cap H_k^{(1)}$ nor $H_j^{(2)} \cap H_k^{(2)}$, the \eqref{fwerex} does not hold and the family-wise error rate is not controlled.
 Let us verify this.
 Now we assume that there is no correlation between $y_{i1}^{(u)}$ and $y_{i2}^{(u)}$, i.e., $\sigma_{12}=0$.
 In addition, we assume that the true hypothesis is $H_1^{(1)} \cap H_2^{(1)} \cap K_1^{(2)} \cap K_2^{(2)}$ and that $\mu_1^{(2)}-\mu_1^{(0)}$ and $\mu_2^{(2)}-\mu_2^{(0)}$ are large enough. 
 Then, the spurious correlation $\hat{\hat{\sigma}}_{12}/\sqrt{\hat{\hat{\sigma}}_{11}\hat{\hat{\sigma}}_{22}}$ becomes close to $1$, and so $\hat{\hat{X}}_1^{(1)}$ and $\hat{\hat{X}}_2^{(1)}$ ($\hat{\hat{X}}_1^{(2)}$ and $\hat{\hat{X}}_2^{(2)}$) are almost the same random variables.
 Therefore, the family-wise error rate is evaluated as
\begin{align*}
&\P_{H_1^{(1)}\cap H_2^{(2)}}\{\max (T_1^{(1)},T_2^{(1)})>c\}
\approx\P\{\max (X_1^{(1)},X_2^{(1)})>c\}
\\&>\P\{\max (\hat{\hat{X}}_1^{(1)},\hat{\hat{X}}_1^{(2)})>c\}
\approx\P\{\max (\hat{\hat{X}}_1^{(1)},\hat{\hat{X}}_2^{(1)},\hat{\hat{X}}_1^{(2)},\hat{\hat{X}}_2^{(2)})>c\}
=\alpha,
\end{align*}
and we can see that it asymptotically exceeds $\alpha$. 
 Here, $(X_1^{(1)},X_2^{(1)})$ is a weak limit of $(T_1^{(1)},T_2^{(1)})$. 
 The inequality holds because of the independence between $X_1^{(1)}$ and $X_2^{(1)}$ and the positivity of the correlation between $\hat{\hat{X}}_1^{(1)}$ and $\hat{\hat{X}}_1^{(2)}$. 

Similar to the example of the three groups above, in the setting of this section, we can consider the use of 
\begin{align}
\hat{\hat{\sigma}}_{jk}^{(u)} \equiv \frac{1}{n^{(u)}-d^{(u)}}\sum_{i=1}^{n^{(u)}} \bigg(y_{ij}^{(u)}-\frac{\sum_{v=0}^m \sum_{i=1}^{n^{(v)}} y_{ij}^{(v)}}{\sum_{v=0}^mn^{(v)}}\bigg) \bigg(y_{ik}^{(u)}-\frac{\sum_{v=0}^m \sum_{i=1}^{n^{(v)}} y_{ik}^{(v)}}{\sum_{v=0}^mn^{(v)}}\bigg)
\label{spur2}
\end{align}
for an appropriately defined $d^{(u)}$ ($u\in\{0,1,\ldots,m\}$); however, this also does not satisfy the control of the family-wise error rate as in the following corollary (See Web Appendix for its proof):

\begin{coro}
The \maxt\ method using $(\hat{\hat{\sigma}}_{jk}^{(0)},\hat{\hat{\sigma}}_{jk}^{(s)})$ for $\rho_{jk}^{(ss)}$ in \eqref{asympcor1} and $\rho_{jk}^{(st)}$ in \eqref{asympcor2} does not asymptotically control the family-wise error rate even when $m\ge 2$.
\label{co1}
\end{coro}


\section{Proposal for spurious correlation}\label{sec3}

From the previous section, the multiplicity is adjusted if we use $(\tilde{\sigma}_{jk}^{(0)}, \tilde{\sigma}_{jk}^{(s)})$ or $(\hat{\sigma}_{\{s\},jk}^{(0)}, \hat{\sigma}_{\{s\},jk}^{(s)})$ in place of $({\sigma}_{jk}^{(0)}, {\sigma}_{jk}^{(s)})$, which means that the \maxt\ method using $(\theta+1) (\hat{\sigma}_{\{s\},jk}^{(0)}, \hat{\sigma}_{\{s\},jk}^{(s)}) - \theta (\tilde{\sigma}_{jk}^{(0)}, \tilde{\sigma}_{jk}^{(s)})$ for $\rho_{jk}^{(ss)}$ in \eqref{asympcor1} asymptotically controls the family-wise error rate for an arbitrary $\theta$.

Let us consider what value should be used as $\theta$.
 When $(\hat{\sigma}_{\{s\},jk}^{(0)}, \hat{\sigma}_{\{s\},jk}^{(s)})$ is assumed to be larger than $(\tilde{\sigma}_{jk}^{(0)}, \tilde{\sigma}_{jk}^{(s)})$, and if we assign a large value to $\theta$, the estimator $(\theta+1) (\hat{\sigma}_{\{s\},jk}^{(0)}, \hat{\sigma}_{\{s\},jk}^{(s)}) - \theta (\tilde{\sigma}_{jk}^{(0)}, \tilde{\sigma}_{jk}^{(s)})$ becomes large and this enhances the statistical power.
 On the other hand, the control of the family-wise error rate becomes unstable even though it is asymptotically controlled because the variance of the estimator becomes large.
 Therefore, we consider setting the value of $\theta$ as large as possible while we assure the stability of the control to some extent.
 Specifically, we propose the use of the supremum of $\theta$ such that the variances of $(\theta+1) (\hat{\sigma}_{\{s\},jk}^{(0)}, \hat{\sigma}_{\{s\},jk}^{(s)}) - \theta (\tilde{\sigma}_{jk}^{(0)}, \tilde{\sigma}_{jk}^{(s)})$ are asymptotically equal to or smaller than those of the conventional estimators $(\tilde{\sigma}_{jk}^{(0)}, \tilde{\sigma}_{jk}^{(s)})$.

Let us asymptotically evaluate the variance of the estimator in a setting of two-group comparisons, i.e. $m=1$, with a common covariance matrix $\bm{\Sigma}^{(0)}=\bm{\Sigma}^{(1)}=\bm{\Sigma}$.
 In asymptotics, we fix $n^{(1)}/n^{(0)}$ and increase $n\equiv n^{(0)}+n^{(1)}$.
 Firstly, by letting $A_{jk} \equiv \sum_{u=0}^1 \sum_{i=1}^{n^{(u)}} (y_{ij}^{(u)}-\bar{y}_j^{(u)}) (y_{ik}^{(u)}-\bar{y}_k^{(u)})$, the variance of the conventional estimator is evaluated as
\begin{align}
{\V} (\tilde{\sigma}_{jk}) = {\V} \bigg( \frac{A_{jk}}{n-2} \bigg) \approx \frac{1}{n}(\sigma_{jj} \sigma_{kk} + \sigma_{jk}^2) + \frac{2}{n^2} (\sigma_{jj} \sigma_{kk} + \sigma_{jk}^2).
\label{veval1}
\end{align}
On the other hand, by letting $B_{jk} \equiv n^{(0)} n^{(1)} (\bar{y}_j^{(0)}-\bar{y}_j^{(1)}) (\bar{y}_k^{(0)}-\bar{y}_k^{(1)}) / n$, which is independent of $A_{jk}$, the variance of the estimator under $H_j^{(1)} \cap H_k^{(1)}$ is evaluated as
\begin{align*}
{\V} (\hat{\sigma}_{jk}) 
 = {\V} \bigg( \frac{A_{jk}}{n-1} \bigg) + {\V} \bigg( \frac{B_{jk}}{n-1} \bigg)
 \approx \frac{1}{n} (\sigma_{jj} \sigma_{kk} + \sigma_{jk}^2) + \frac{1}{n^2} \sigma_{jj} \sigma_{kk}.
\end{align*}
 Because $(\theta+1)\hat{\sigma}_{jk}-\theta\tilde{\sigma}_{jk}$ is written as $\{(\theta+1)/(n-1) - \theta/(n-2)\}A_{jk} + \{(\theta+1)/(n-1)\}B_{jk}$, its variance is evaluated as
\begin{align}
{\V} \{(\theta+1)\hat{\sigma}_{jk}-\theta\tilde{\sigma}_{jk}\} \approx \frac{1}{n} (\sigma_{jj} \sigma_{kk} + \sigma_{jk}^2) + \frac{1}{n^2} \{(\theta^2+1) \sigma_{jj} \sigma_{kk} - 2\theta\sigma_{jk}^2\}.
\label{veval2}
\end{align}
 The difference between the right sides of \eqref{veval1} and \eqref{veval2} is $\{2(\theta+1)\sigma_{jk}^2 + (1-\theta^2)\sigma_{jj}\sigma_{kk}\}/n^2$, and we can see that ${\V} \{(\theta+1)\hat{\sigma}_{jk}-\theta\tilde{\sigma}_{jk}\}$ is always asymptotically smaller than ${\V} (\tilde{\sigma}_{jk})$ when $-1<\theta<1$.
For more details, see the Appendix, which derives the following theorem in a similar way.

Moreover, in the setting of the general theory in Section \ref{sec2}, a similar property holds under the following two requirements: the sample sizes in the groups are close to each other; if the variance of a variable is large, the variance of the other variables is also large. This property is written in the following theorem.
 Therefore, we propose that $\theta=1$ in this paper.

\begin{theo}
Let us assume that $(\sigma_{jj}^{(s)}-\sigma_{jj}^{(0)}) (\sigma_{kk}^{(s)}-\sigma_{kk}^{(0)}) > 0$ in the setting of Section \ref{sec2} ($1\le s\le m$, $1\le j<k\le p$). 
When $n^{(0)}$ and $n^{(s)}$ are close enough, the supremum of $\theta$, such that the variance of $(\theta+1) \hat{\sigma}_{\{s\},jk}^{(u)} - \theta \tilde{\sigma}_{jk}^{(u)}$ is always smaller than that of $\tilde{\sigma}_{jk}$, is $1$.
\label{th2}
\end{theo}


\section{Simulation study}\label{sec4}

Let us compare the performances of the proposed and existing methods through simulation studies in a simple setting.
 Considering the real data analysis in the next section, we treat a two-group comparison, i.e. $m=1$, and assume the sample size $n$ in each group to be between $6$ to $18$ and the number of tests $p$ on each group to be between $20$ to $80$.
 In addition, we assume that the covariance matrix is a block-diagonal matrix whose block is a $10\times 10$ uniform covariance matrix with variance $1$ and covariance $\rho$, and $rho$ is assumed to be between $0.0$ and $0.6$.
 Letting $r$ be the rate of the true alternative hypotheses in all alternative hypotheses, we assume that the differences between the expectations are $\mu$ in the true alternative hypotheses.
 Among existing methods, we consider the Bonferroni, the \maxt\, and the step-down \maxt\ methods. The step-down \maxt\ method uses the \maxt\ method in a step-down procedure (\citealt{DudLaa07}).
 It is trivial that we can use our method in a step-down procedure, and we refer to this as ``Proposal.''

By denoting the estimates of the correlation matrix for $\{T_j^{(s)}\ |\ 1\le j\le p,\ 1\le s\le m\}$ made by the proposal in Section \ref{sec3} as $\hat{\bm{\Psi}}$, in some cases, $\hat{\bm{\Psi}}$ does not become positive-definite and our method cannot be applied.
 In such cases, letting $\tilde{\bm{\Psi}}$ be the estimates of the correlation matrix for $\{T_j^{(s)}\ |\ 1\le j\le p,\ 1\le s\le m\}$ made by $\rho_{jk}^{(ss)}$ in \eqref{asympcor1}, $\rho_{jk}^{(st)}$ in \eqref{asympcor2}, and $\rho_{jj}^{(st)}$ in \eqref{asympcor3}, we gradually move $\tilde{\bm{\Psi}}$ closer to $\hat{\bm{\Psi}}$ as long as the positive-definiteness is maintained, and we use the last matrix before the positive-definiteness is broken. 
 We specifically provide an output of the following algorithm:

\begin{algo}
Positive-definitization by increasing values of components.
\begin{list}{}{\setlength{\labelwidth}{5mm}\setlength{\leftmargin}{5mm}}
\item[i.]
Set $\bm{\Psi}=\tilde{\bm{\Psi}}$.
\item[ii.]
Randomly select an element $\psi$ from $\bm{\Psi}$, and select the corresponding element $\hat{\psi}$ from $\hat{\bm{\Psi}}$. 
\item[iii.]
Replace $\psi$ with $\psi + 0.2(\hat{\psi}-\psi)$ if the replaced $\bm{\Psi}$ is positive-definite.  
\item[iv.]
Repeat ii and iii as long as an update exists, and when the update does not occur, output $\bm{\Psi}$.
\end{list}
\label{algo2}
\end{algo}

Firstly, we will check how the proposed method, which asymptotically controls the family-wise error rate, controls the family-wise error rate in finite sample cases.
 Table \ref{tab1} numerically evaluates the family-wise error rate by each method when the significance level is $5\%$.
 It is to be noted that we include settings in which there are true alternative hypotheses to verify the differences of the proposed and existing methods even though the family-wise error rate for the proposed method is clearly smaller than $5\%$ in these settings. 
 From the table, it can be seen that the proposed and existing methods share almost the same values under complete null hypotheses; that is, the proposed method controls the family-wise error rate accurately enough even if the sample size is not large.
 When there are true alternative hypotheses, the family-wise error rate in the proposed method is closer to $5\%$ than that in the existing methods. 
 This indicates that the proposed method is superior to the existing methods in terms of statistical power.

\begin{table}
\def~{\hphantom{0}}
\caption{Family-wise error rates for proposed and existing methods. The values without and in parentheses are respectively averages of family-wise error rates and their standard deviations ({\%}) with significance level $5\%$.}
\begin{center}
\begin{tabular}{ccccccccc}
$\rho$ & $n$ & $p$ & $\mu$ & $r$ & & \textsc{Bon} & \textsc{SD\maxt} & \textsc{Proposal} \\
0.0 & 12 & 50 & -- & 0.0 & & 4.53 & 4.64 \ \ (0.54) & 4.64 \ \ (0.56) \\
0.2 & 12 & 50 & -- & 0.0 & & 4.57 & 4.82 \ \ (0.64) & 4.80 \ \ (0.64) \\
0.4 & 12 & 50 & -- & 0.0 & & 4.28 & 4.91 \ \ (0.67) & 5.01 \ \ (0.81) \\
0.6 & 12 & 50 & -- & 0.0 & & 3.77 & 5.07 \ \ (0.82) & 5.26 \ \ (1.20) \\
0.3 & ~6 & 50 & -- & 0.0 & & 3.53 & 4.38 \ \ (0.75) & 4.21 \ \ (0.82) \\
0.3 & 10 & 50 & -- & 0.0 & & 3.74 & 4.29 \ \ (0.69) & 4.32 \ \ (0.71) \\
0.3 & 14 & 50 & -- & 0.0 & & 4.29 & 4.77 \ \ (0.70) & 4.78 \ \ (0.74) \\
0.3 & 18 & 50 & -- & 0.0 & & 4.44 & 4.90 \ \ (0.72) & 4.84 \ \ (0.71) \\
0.3 & 12 & 20 & -- & 0.0 & & 4.26 & 4.92 \ \ (0.53) & 4.96 \ \ (0.55) \\
0.3 & 12 & 40 & -- & 0.0 & & 4.37 & 4.91 \ \ (0.62) & 4.88 \ \ (0.67) \\
0.3 & 12 & 60 & -- & 0.0 & & 4.46 & 4.93 \ \ (0.72) & 4.86 \ \ (0.71) \\
0.3 & 12 & 80 & -- & 0.0 & & 4.59 & 5.04 \ \ (0.82) & 4.99 \ \ (0.83) \\
0.3 & 12 & 50 & 0.6 & 0.5 & & 2.39 & 2.66 \ \ (0.40) & 2.81 \ \ (0.49) \\
0.3 & 12 & 50 & 1.0 & 0.5 & & 2.39 & 2.67 \ \ (0.37) & 3.06 \ \ (0.58) \\
0.3 & 12 & 50 & 1.4 & 0.5 & & 2.39 & 2.65 \ \ (0.39) & 3.27 \ \ (0.54) \\
0.3 & 12 & 50 & 1.8 & 0.5 & & 2.39 & 2.68 \ \ (0.40) & 3.40 \ \ (0.57) \\
0.3 & 12 & 50 & 1.2 & 0.2 & & 3.60 & 4.09 \ \ (0.69) & 4.19 \ \ (0.71) \\
0.3 & 12 & 50 & 1.2 & 0.4 & & 2.82 & 3.11 \ \ (0.41) & 3.53 \ \ (0.62) \\
0.3 & 12 & 50 & 1.2 & 0.6 & & 1.96 & 2.28 \ \ (0.37) & 2.85 \ \ (0.52) \\
0.3 & 12 & 50 & 1.2 & 0.8 & & 1.05 & 1.18 \ \ (0.19) & 1.78 \ \ (0.43) \\
\end{tabular}
\end{center}
\textsc{Bon}, Bonferroni method; \textsc{SD\maxt}, step-down \maxt\ method.
\label{tab1}
\end{table}

Next, we will verify the superiority of the proposed method.
 Letting $\alpha=5\%$ and $r=1.0$, Table \ref{tab2} numerically evaluates the statistical power of each method. 
 We observe the degrees of improvements by considering correlations and by using a step-down procedure from the difference between the Bonferroni and the \maxt\ methods and from the difference between the \maxt\ and the step-down \maxt\ methods, respectively.
 We would like to state that such improvements are sometimes overwhelmed by the improvement of the proposed method when compared to the step-down \maxt\ method.
 Especially when the correlation $\rho$, the number of tests $p$, and the difference between the expectations $\mu$ are large, our method significantly increases statistical power.

\begin{table}
\def~{\hphantom{0}}
\caption{Statistical power of proposed and existing methods. The values without and within brackets are the powers and averages of $p$-values ({\%}), respectively.}
\begin{center}
\begin{tabular}{ccccccccc}
$\rho$ & $n$ & $p$ & $\mu$ & & \textsc{Bon} & \maxt & \textsc{SD\maxt} & \textsc{Proposal} \\
0 & 12 & 50 & 1.2 & & 30.4 \ \ [34.4] & 30.9 \ \ [33.8] & 34.7 \ \ [28.8] & 38.8 \ \ [23.7] \\
0.2 & 12 & 50 & 1.2 & & 33.0 \ \ [31.9] & 34.1 \ \ [29.6] & 39.7 \ \ [25.4] & 47.2 \ \ [19.7] \\
0.4 & 12 & 50 & 1.2 & & 33.2 \ \ [32.2] & 35.5 \ \ [27.5] & 41.7 \ \ [24.3] & 51.5 \ \ [18.2] \\
0.6 & 12 & 50 & 1.2 & & 30.5 \ \ [34.1] & 36.3 \ \ [25.1] & 41.3 \ \ [23.3] & 54.6 \ \ [16.5] \\
0.3 & ~6 & 50 & 1.2 & & ~6.9 \ \ [65.9] & ~7.7 \ \ [59.6] & ~8.2 \ \ [58.6] & 12.4 \ \ [50.3] \\
0.3 & 10 & 50 & 1.2 & & 21.2 \ \ [41.8] & 23.1 \ \ [37.4] & 25.9 \ \ [34.4] & 35.6 \ \ [25.9] \\
0.3 & 14 & 50 & 1.2 & & 40.8 \ \ [24.8] & 42.1 \ \ [21.9] & 50.2 \ \ [18.1] & 57.3 \ \ [13.8] \\
0.3 & 18 & 50 & 1.2 & & 59.8 \ \ [14.1] & 61.6 \ \ [12.5] & 70.2 \ \ ~[8.7] & 74.3 \ \ ~[7.0] \\
0.3 & 12 & 20 & 1.2 & & 46.8 \ \ [20.4] & 48.6 \ \ [17.1] & 57.1 \ \ [13.4] & 64.0 \ \ [10.3] \\
0.3 & 12 & 40 & 1.2 & & 35.4 \ \ [29.0] & 36.7 \ \ [25.4] & 43.1 \ \ [21.7] & 52.1 \ \ [16.3] \\
0.3 & 12 & 60 & 1.2 & & 29.9 \ \ [35.0] & 31.1 \ \ [31.4] & 35.7 \ \ [27.9] & 44.2 \ \ [21.1] \\
0.3 & 12 & 80 & 1.2 & & 26.3 \ \ [39.9] & 27.3 \ \ [35.6] & 31.8 \ \ [32.2] & 41.6 \ \ [24.7] \\
0.3 & 12 & 50 & 0.9 & & 14.3 \ \ [55.0] & 14.9 \ \ [50.6] & 16.9 \ \ [48.5] & 21.8 \ \ [42.3] \\
0.3 & 12 & 50 & 1.1 & & 23.4 \ \ [40.8] & 24.6 \ \ [36.5] & 28.8 \ \ [33.3] & 36.7 \ \ [26.4] \\
0.3 & 12 & 50 & 1.3 & & 43.3 \ \ [23.8] & 45.3 \ \ [21.0] & 52.2 \ \ [16.9] & 60.5 \ \ [12.1] \\
0.3 & 12 & 50 & 1.5 & & 55.8 \ \ [14.9] & 57.8 \ \ [12.9] & 67.8 \ \ ~[8.8] & 77.1 \ \ ~[5.6]
\end{tabular}
\end{center}
\textsc{Bon}, Bonferroni method; \textsc{SD\maxt}, step-down \maxt\ method.
\label{tab2}
\end{table}


\bibliography{List}

\end{document}